# CO₂ MITIGATION MODEL FOR CHINA'S RESIDENTIAL BUILDING SECTOR

Minda Ma [1*, 2] (maminda@cqu.edu.cn) , Weiguang Cai [1*, 2, 3] (wgcai@cqu.edu.cn)

1. International Research Center for Sustainable Built Environment, Chongqing University, 400045, PR China

2. Special Committee of Building Energy Consumption Statistics, China Association of Building Energy Efficiency, Beijing, 100835, PR China

3. Energy Technologies Area, Lawrence Berkeley National Laboratory, Berkeley, CA 94720, USA

**Abstract**

This paper aims to investigate the factors that can mitigate carbon-dioxide ($CO_2$) intensity and further assess CMRBS in China based on a household scale via decomposition analysis. Here we show that: Three types of housing economic indicators and the final emission factor significantly contributed to the decrease in $CO_2$ intensity in the residential building sector. In addition, the CMRBS from 2001–2016 was 1816.99 $MtCO_2$, and the average mitigation intensity during this period was 266.12 $kgCO_2 \cdot (household \cdot year)^{-1}$. Furthermore, the energy-conservation and emission-mitigation strategy caused CMRBS to effectively increase and is the key to promoting a more significant emission mitigation in the future. Overall, this paper covers the CMRBS assessment gap in China, and the proposed assessment model can be regarded as a reference for other countries and cities for measuring the retrospective $CO_2$ mitigation effect in residential buildings.

**Abbreviation notation**
CMRBS – Carbon mitigation in residential building sector
ECEM – Energy-conservation and emission-mitigation
FYP – Five-Year Plan
LMDI – Log-Mean Divisia index
Mtce – Million tons of standard coal equivalent
$MtCO_2$ – Million tons of carbon dioxide
**Nomenclature**
$C$ – $CO_2$ released from residential buildings
$c$ – $CO_2$ emission per floor space in residential buildings
$c_h$ – $CO_2$ emission per household in residential buildings (i.e., $CO_2$ intensity in residential buildings)
$d$ – Housing purchasing power
$E$ – Energy consumption in residential buildings
$e$ – Energy consumption per floor space in residential buildings
$F$ – Floor spaces of residential buildings
$H$ – Amount of households
$I$ – Income of households
$i$ – Per capita income
$K$ – Final emission factor of residential buildings
$P$ – Population size
$p$ – Population size per household
$P_r$ – Housing price
$r$ – Housing price-to-income ratio
$S$ – Household age structure
**Greek letter**
$\Delta c_h|_{0 \to T}$ – $c_h$ changes during Period $\Delta T$
$\Delta c_{h_d}$ – Effect of $d$ on $c_h$
$\Delta c_{h_e}$ – Effect of $e$ on $c_h$
$\Delta c_{h_i}$ – Effect of $i$ on $c_h$
$\Delta c_{h_K}$ – Effect of $K$ on $c_h$
$\Delta c_{h_p}$ – Effect of $p$ on $c_h$
$\Delta c_{h_r}$ – Effect of $r$ on $c_h$
$\Delta c_{h_S}$ – Effect of $S$ on $c_h$

## 1. Introduction

IPCC [1] has declared that effective mitigation measures for the carbon-dioxide ($CO_2$) emissions in the residential building sector are significant to suppress critical global warming trends [1] since residential buildings are responsible for nearly 20% of the final energy demand, which causes over 22% of $CO_2$ emissions worldwide [2]. Regarding the developing country such as China, its residential building sector is facing a growing demand for household energy service. Therefore, large amounts of primary energy (e.g., coal and natural gas) and secondary energy (e.g., electrical power) have been consumed, which leads to large emission of $CO_2$ in residential buildings [3]. It has been reported that the $CO_2$ which is released from the residential building sector has grown rapidly with a 6.57% increase per year over the past decade in China, and emissions measured at over 1.2 billion tons of $CO_2$ in 2016 [4, 5].

---
[1] Intergovernmental Panel on Climate Change (IPCC).

Reducing the large $CO_2$ emissions from the residential building sector is critical for China in achieving its 2030 emission peak target, many scholars agree that the emission mitigation potential from residential buildings is considerable. As a typical case, McNeil et al. forecasted that effective mitigation measures for the $CO_2$ emissions of residential buildings will contribute approximately 30% to the 2030 emission peak target in China [6], and this viewpoint has been further verified by their latest study [7]. Tan et al. assessed the contribution of the $CO_2$ mitigation potential to emission peak goals in the Chinese residential building sector, and they believed that the turning point of the emission peak will appear before 2030 (approximately in 2027) in the best emission mitigation scenario [8]. **Nevertheless**, regarding the retrospective $CO_2$ mitigation volume, it has been barely investigated and assessed, especially concerning the mitigation indicator of $CO_2$ intensity, which is required to be preferentially analyzed in the emission peak scenario [9]. **Therefore, three questions are proposed for the Chinese residential building sector, as shown below.**

- Q1: Are the intensity and total values of $CO_2$ emissions mitigated in the retrospective phase?
- Q2: What leads to emission mitigation if it does exist?
- Q3: How should the future mitigation effect be strengthened to achieve China's 2030 emission peak?

**To answer the questions above, this paper first** investigates the factors that can mitigate $CO_2$ intensity and further assesses the $CO_2$ mitigation in residential building sector (i.e., CMRBS, which includes both the intensity and total values) of China from 2000–2016 via the decomposition analysis. Furthermore, the strategy for the energy-conservation and emission-mitigation (ECEM) of residential buildings is retrospected to explore policy patterns to achieve more emission mitigation effects in the future.

**The most significant contribution of this paper is** to assess both the intensity and total values of CMRBS based on a household scale. To date, no studies on such a topic have been performed in China to the best of authors' knowledge. Some of the recent literature have already reported cases on $CO_2$ mitigation assessment of the building sector. However, their target is primarily focused on the $CO_2$ mitigation of commercial buildings [10, 11], or their assessment model follows the up-bottom approach which has yet to consider the effect of household scale on CMRBS and only focuses on the assessment level of energy savings instead of $CO_2$ mitigation [12, 13].

The remainder of this paper is conducted as follows: Section 2 presents the materials and methods. Section 3 illustrates the decomposition result of the $CO_2$ intensity and the assessment result of CMRBS. Section 4 discusses the ECEM strategy of the residential building sector and relevant policy implications are put forward. Section 5 focuses on conclusions and upcoming studies.

## 2. Materials and Methods

It is widely accepted that the smallest unit of energy survey in a residential building is the household scale, and $CO_2$ intensity is usually characterized as the $CO_2$ emission per household [14]. Thus, $CO_2$ intensity feature should first be analyzed for conducting the investigation of the factors affecting $CO_2$ mitigation and the assessment of $CO_2$ mitigation. Decomposition analysis has been widely adopted to investigate the factors that can mitigate $CO_2$ intensity and to assess the $CO_2$ mitigation in emission sectors [15]. Regarding the residential building sector of China, this paper presented the CMRBS assessment via a decomposition analysis. To achieve this goal, Kaya identity [16] on $CO_2$ intensity is required to be built as the first step in deploying the decomposition analysis. To present a reasonable decomposition process and ensure the reliability of the decomposition results, this paper referred to the latest highly related study: Liang et al. reported a case of Kaya identity on $CO_2$ intensity in the residential building sector, which considered social, economic and technical features of residential buildings [17], as illustrated in Fig. 1.

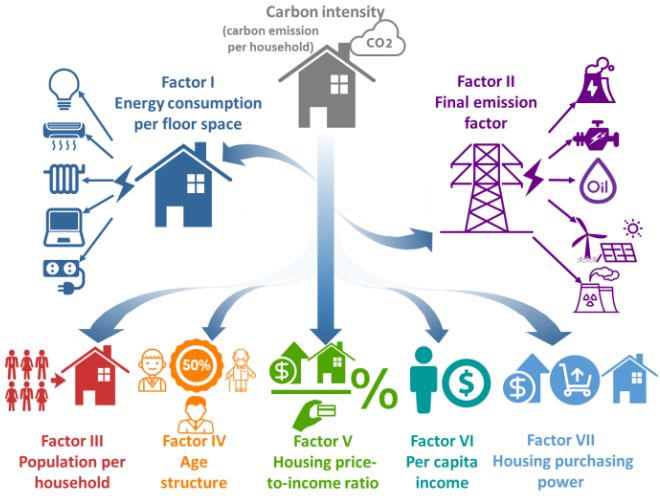

**Fig. 1.** Decomposition flow of Kaya identity on $CO_2$ intensity in the residential building sector.

Mathematical expression of Fig. 1 is as follows.

$$\frac{C}{H} = c_h = \frac{P}{H} \cdot \sum S_j \cdot \frac{\frac{F \cdot P_r}{H}}{\frac{P \cdot I}{H \cdot P}} \cdot \frac{I}{P} \cdot \frac{1}{P_r} \cdot \frac{E}{F} \cdot \sum K_l \quad (1)$$

Let $c_h = \frac{C}{H}$, $p = \frac{P}{H}$, $S = \sum S_j$, $r = \frac{\frac{F \cdot P_r}{H}}{\frac{P \cdot I}{H \cdot P}}$, $i = \frac{I}{P}$, $d = \frac{1}{P_r}$, $e = \frac{E}{F}$, and $K = \sum K_l$; Eq. 1 can be converted to Eq. 2.

$$c_h = p \cdot S \cdot r \cdot i \cdot d \cdot e \cdot K \quad (2)$$

Thereafter, this paper used the Log-Mean Divisia index (LMDI) [18] to decompose Eq. 2 to confirm the effects of seven factors on $CO_2$ intensity. LMDI has been widely applied with Kaya identity to evaluate the effects of various factors on energy consumption or $CO_2$ emissions [19, 20]. Through the LMDI application handbook [21], the changes of $CO_2$ intensity in residential buildings during Period $\Delta T$ ($\Delta c_h|_{0 \to T}$) are decomposed as follows:

$$\Delta c_h|_{0 \to T} = c_h|_T - c_h|_0 = \Delta c_{h_p} + \Delta c_{h_S} + \Delta c_{h_r} + \Delta c_{h_i} + \Delta c_{h_d} + \Delta c_{h_e} + \Delta c_{h_K} \quad (3)$$

Specifically, every parameter (e.g., $\Delta c_{h_p}$) on the right side of Eq. 3 can be further expressed as:

$$\Delta c_{h_p} = L(c_h|_T, c_h|_0) \cdot \ln\left(\frac{p|_T}{p|_0}\right) = L(c_h|_T, c_h|_0) \cdot \ln\left(\frac{P|_T \cdot H|_0}{P|_0 \cdot H|_T}\right) \quad (4)$$

where $L(a,b) = \begin{cases} \frac{a-b}{\ln a - \ln b}, & a \neq b \ (a > 0, \ b > 0) \\ 0, & a = b \ (a > 0, \ b > 0) \end{cases} \quad (5)$

Furthermore, CMRBS is expressed as follows:

$$\text{CMRBS intensity}|_{0 \to T} = \sum |\Delta c_{h_m}|_{0 \to T}| \quad (6)$$

$$\text{CMRBS}|_{0 \to T} = H|_{0 \to T} \times \left(\sum |\Delta c_{h_m}|_{0 \to T}|\right) \quad (7)$$

where $\Delta c_{h_m}|_{0 \to T} \in \{\Delta c_{h_p}, \Delta c_{h_S}, \Delta c_{h_r}, \Delta c_{h_i}, \Delta c_{h_d}, \Delta c_{h_e}, \Delta c_{h_K}\}$, $\Delta c_{h_m}|_{0 \to T} < 0 \quad (8)$

## 3. Results

### 3.1. Decomposition results on $CO_2$ intensity in the residential building sector

Fig. 2 presents the LMDI decomposition results of $CO_2$ intensity in the Chinese residential building sector during the 2000-2016 period via Eqs. 3 to 5. For the positive factors promoting the $CO_2$ intensity growth in the residential building sector, per capita income played the most significant role, as illustrated by the light green blocks shown in Fig. 2. Compared to the per capita income, energy consumption per floor space in the residential building sector also contributed positively to the growth of $CO_2$ intensity during 2000–2016, as expressed by the blue blocks shown in Fig. 2. This finding reveals the significant coupling of energy and emission intensities, which is linked by the final emission factor of the residential building sector (the purple blocks shown in Fig. 2).

Regarding the negative factors, the housing purchasing power contributed the most in decreasing the $CO_2$ intensity, as demonstrated by the light blue blocks shown in Fig. 2. Compared to the housing purchasing power, the housing price-to-income ratio also contributed negatively in terms of increasing the $CO_2$ intensity from 2000–2008, as expressed by the green blocks shown in Fig. 2. However, the impact of the housing price-to-income ratio on $CO_2$ intensity shifted from a negative status to a positive status from 2008–2016.

As indicated in Eq. 1, the household age structure and the final emission factor in the residential building sector can be further extended into a series of subfactors (i.e., $S = \sum S_j$ and $K = \sum K_l$), respectively. Hence, the red and purple blocks shown in Fig. 2 were further decomposed. Regarding the energy structure changes, an optimal energy structure can lead to the maximum potential of $CO_2$ mitigation. From 2000–2016, the proportion of coal consumption in the residential building sector decreased significantly (from 44.33% to 26.18%), and the proportions of electricity and heating increased during the same period (from 40.57% to 50.73%). Let the 2012–2016 period serve an example, where the $CO_2$ mitigation contributions of the five main energy sources are summarized as follows: 69.60 (coal), 28.64 (oil), 32.77 (natural gas), 65.93 (electricity), and 68.96 (heating) $kgCO_2$ per household. Regarding the impacts of population size per household and the household age structure on $CO_2$ intensity, the residents within the 15–64 age group were the main force in promoting the $CO_2$ intensity decrease from 2000–2012. However, the impact of the population size per household on the $CO_2$ intensity shifted from a negative status to a positive status from 2012–2016. Thus, the contribution of residents within the 15–64 age group to the $CO_2$ intensity increase changed from a negative status to a positive status from 2012–2016.

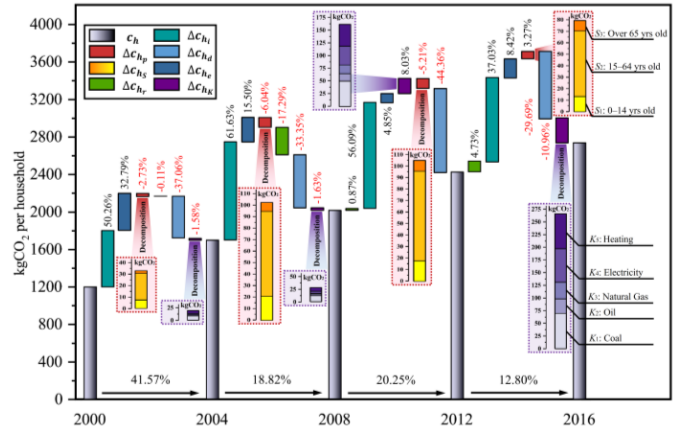

**Fig. 2.** Changes of $CO_2$ intensity in the Chinese residential building sector ($\Delta c_h$) via a decomposition analysis (2000–2016).

*3.2. Retrospective $CO_2$ mitigation in the residential building sector*

Fig. 3 a reflects the trends on total and intensity values of CMRBS from 2001–2016 in China via the calculation based on Eqs. 6 to 8. To express the uncertainty level of CMRBS, two error bands were added in Fig. 3 a [error band value of CMRBS intensity: 89.45 kgCO₂ · (household · year)⁻¹, and error band value of CMRBS: 40.19 million tons of carbon dioxide (MtCO₂) per year]. Furthermore, the average intensity of CMRBS during different Five-year Plan (FYP) periods in China is summarized as follows: 199.75 (10th FYP Period: 2001–2005), 307.19 (11th FYP Period: 2006–2010), and 284.45 kgCO₂· (household · year)⁻¹ (12th FYP Period: 2011–2015). Moreover, CMRBS values during the three periods above are: 393.68 (10th FYP Period), 648.10 (11th FYP Period), and 641.40 MtCO₂ (12th FYP Period). Besides, Fig. 3 c assessed CMRBS intensity at another two scales (i.e., $CO_2$ mitigation per capita and $CO_2$ mitigation per floor space). The fitting estimations shown in Fig. 3 b and d reflect that the continuous growth of CMRBS at four different scales is obvious.

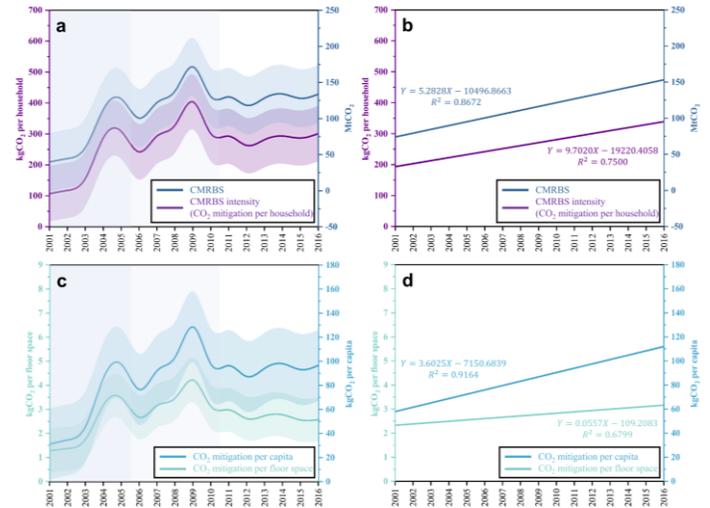

**Fig. 3 a and b.** Total and intensity values of CMRBS in China during 2001–2016; **c and d.** CMRBS intensity at another two scales during 2001–2016 ($CO_2$ mitigation per capita and $CO_2$ mitigation per floor space).

After assessing CMRBS values, a comparative analysis of the official expected and assessed values of energy savings in the residential building sector was presented, as illustrated in Fig. 4. To provide a comparable condition, the CMRBS values of Fig. 3 were converted to the energy-saving values. Fig. 4 reveals that the assessed values were much higher than the official expected values during the 11th and 12th FYP Periods. Furthermore, although the 13th FYP Period (2016–2020) is still in process, the assessed energy-saving value in 2016 reached 60.80 million tons of standard coal equivalent (Mtce), constituting over 60% of the official expected value from 2016–2020. It is believed that the residential building sector is able to meet its 13th FYP energy-saving target. It should be noted that the Chinese government has officially conducted the nationwide effort to reach the ECEM target since 2006, which means that the official expected value of energy savings in the residential building sector during the 10th FYP Period (2001–2005) was lacking and the relevant comparative analysis of energy savings is difficult to represent, as is shown in Fig. 4. However, the results of the comparative analysis focusing on 2006–2016 are enough to prove that the Chinese residential building sector has achieved obvious energy-saving benefits over the past decade.

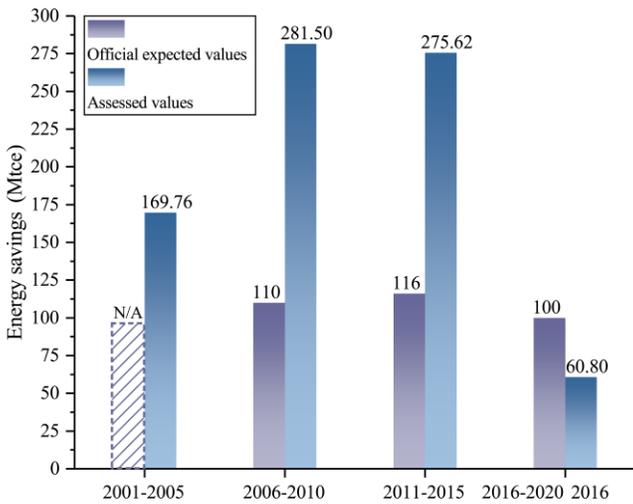

**Fig. 4.** Assessed and official expected values of energy savings in the Chinese residential building sector.

## 4. Discussion

ECEM strategy needs to be reviewed for the root cause promoting the increase of CMRBS. ECEM strategy of the building sector is defined as the set of code, act, policy document, energy conservation standard, energy efficiency label, energy conservation technology, and economic intensive approach to achieve the ECEM target in the building sector. Specifically, ECEM strategy can be mainly summarized into three parts: a. mandatory strategy (e.g., energy conservation standard), b. information strategy (e.g., energy efficiency label, ladder-type price of electricity), and c. economic intensive strategy (e.g., special funding, financial subsidy) [22]. Regarding the residential building sector of China, its official ECEM strategy was fully deployed at the beginning of 11[th] FYP Period (2006), which includes over 10 relevant codes and acts, more than 80 policy documents, and at least 50 mandatory standards by the end of 2015. For example, to fully promote the ECEM strategy in the building sector, the Chinese government issued *China Energy Conservation Code* (1997, 2007, 2016, 2018 versions) and *China Act on Energy Conservation of Civil Buildings* (2008). Following the two codes' guidance above, a series of specific approaches such as the energy conservation standards of newly built buildings, economic intensive approaches on the energy conservation retrofit of existing buildings, the evaluation system of green buildings and the ECEM technology of civil buildings have been proposed for the public, as is mainly illustrated in Fig. 5.

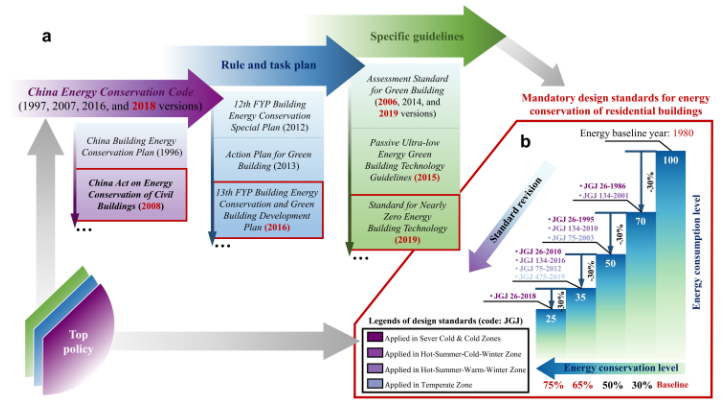

**Fig. 5.** Development pattern of ECEM strategy in the Chinese residential building sector (2006–2019). **Note:** a. principal line of ECEM strategy; b. mandatory design standards for energy conservation of residential buildings across different building climate zones.

According to Fig. 5, it is obviously observed that the Chinese government has expended much effort to develop the ECEM cause, and this significant achievement directly leads to the CMRBS increase. In addition, it is believed that more significant emission mitigation will be realized via the further deployment and implementation of ECEM strategy in the residential building sector in the upcoming phase.

Implications for ECEM strategy in the residential building sector of China are conducted briefly as follows: a. implementing the official policy assessment system for ECEM which is built based on the ECEM effect; b. establishing the multidimensional data statistical system for energy and emission which includes the large sample of building microdata and the sustainable monthly/quarterly/annual data on energy and emission at national and provincial scales; c. issuing the design/construction mandatory standard for residential buildings guided by the ECEM effect; d. increasing the financial expenditure on the formulation and implementation of the ECEM strategy; e. spreading a series of high energy-efficiency technologies and productions such as the microgrid and distributed energy system, and the nearly zero energy building technology.

## 5. Conclusions

This paper investigated the factors for mitigating $CO_2$ intensity and further assessed the intensity and total values of CMRBS in China from 2000–2016 via decomposition analysis. Furthermore, the ECEM strategy of residential buildings was retrospected for exploring policy patterns to achieve more emission mitigation effects in the future.

Upcoming studies need to be deployed to fill in several gaps in the current study. Regarding the factors affecting CMRBS, the impact of climate change on energy and emission in residential buildings is significant. Specifically, the temperature change does affect the energy and emission intensities of residential buildings since residents have different requirements on heating, ventilating and air conditioning system operation in different temperatures. Hence, the future study should attempt to consider the temperature impact on CMRBS. For the unusual emission mitigation peak which occurred in 2009, the internal cause needs to be investigated. Since the time point on the peak is close to the time point of the 2008 financial crisis, the econometrics approach (e.g., regression discontinuity) may be adopted to investigate the cause leading to the peak effect. Furthermore, regarding the case area, since the $CO_2$ emission feature of the residential building sector in urban and rural regions differs significantly, individual studies on provincial-level/city-level $CO_2$ mitigation assessments in residential buildings in urban and rural China are worthy to be conducted, respectively, which will help the central and local governments implement more effective targets and strategies for ECEM. At last, to conduct the sustainable development of built environment, the scope of emission mitigation assessment in built environment needs to be further extended, such as greenhouse gas mitigation, particulate matter mitigation, etc.

**Acknowledgement**

This study was supported by the Fundamental Research Funds for the Central Universities of PR China (106112017CDJXSYY0001-KJYF201706), the National Planning Office of Philosophy and Social Science Foundation of China (19BJY065), and the National Key R&D Program of China (2018YFD1100201).